\documentstyle[epsfig,psfig,sprocl]{article}
\begin{document}

\title{GAMMA-GAMMA, GAMMA-ELECTRON COLLIDERS: \\ 
  ACCELERATOR, LASER  AND INTERACTION REGION ISSUES.}

\author{ V.I. TELNOV }

\address{Institute of Nuclear Physics, 630090 Novosibirsk,
  Russia \\ email:telnov@inp.nsk.su }

\maketitle

\newcommand{\M}{\mbox{m}}
\newcommand{\n}{\mbox{$n_f$}}
\newcommand{\EE}{\mbox{ee}}
\newcommand{\EP}{\mbox{e$^+$}}
\newcommand{\EM}{\mbox{e$^-$}}
\newcommand{\EPEM}{\mbox{e$^+$e$^-$}}
\newcommand{\EMEM}{\mbox{e$^-$e$^-$}}
\newcommand{\GG}{\mbox{$\gamma\gamma$}}
\newcommand{\GE}{\mbox{$\gamma$e}}
\newcommand{\GP}{\mbox{$\gamma$e$^+$}}
\newcommand{\TEV}{\mbox{TeV}}
\newcommand{\GEV}{\mbox{GeV}}
\newcommand{\LGG}{\mbox{$L_{\gamma\gamma}$}}
\newcommand{\LGE}{\mbox{$L_{\gamma e}$}}
\newcommand{\LEE}{\mbox{$L_{ee}$}}
\newcommand{\WGG}{\mbox{$W_{\gamma\gamma}$}}
\newcommand{\EV}{\mbox{eV}}
\newcommand{\CM}{\mbox{cm}}
\newcommand{\MM}{\mbox{mm}}
\newcommand{\NM}{\mbox{nm}}
\newcommand{\MKM}{\mbox{$\mu$m}}
\newcommand{\SEC}{\mbox{s}}
\newcommand{\CMS}{\mbox{cm$^{-2}$s$^{-1}$}}
\newcommand{\MRAD}{\mbox{mrad}}
\newcommand{\IND}{\hspace*{\parindent}}
\newcommand{\E}{\mbox{$\epsilon$}}
\newcommand{\EN}{\mbox{$\epsilon_n$}}
\newcommand{\EI}{\mbox{$\epsilon_i$}}
\newcommand{\ENI}{\mbox{$\epsilon_{ni}$}}
\newcommand{\ENX}{\mbox{$\epsilon_{nx}$}}
\newcommand{\ENY}{\mbox{$\epsilon_{ny}$}}
\newcommand{\EX}{\mbox{$\epsilon_x$}}
\newcommand{\EY}{\mbox{$\epsilon_y$}}
\newcommand{\BI}{\mbox{$\beta_i$}}
\newcommand{\BX}{\mbox{$\beta_x$}}
\newcommand{\BY}{\mbox{$\beta_y$}}
\newcommand{\SX}{\mbox{$\sigma_x$}}
\newcommand{\SY}{\mbox{$\sigma_y$}}
\newcommand{\SZ}{\mbox{$\sigma_z$}}
\newcommand{\SI}{\mbox{$\sigma_i$}}
\newcommand{\SIP}{\mbox{$\sigma_i^{\prime}$}}

\abstracts {In this report on Photon Colliders the following technical
  aspects are considered: special requirements to an accelerator,
  new ideas on laser optics, laser cooling, and interaction region
  layout issues. In fact it is continuation of my first talk at this workshop
  where physics motivation, possible luminosities and backgrounds were
  discussed.}

\section {Introduction}

As a general introduction see my first report from this workshop and
references therein.~\cite{T1}
Photon Colliders are based on \EPEM\ colliders and the main problem is the
same: production of electron beams with low emittances and acceleration to high
energies.  However, photon colliders have several new features and
differences which require special study, especially if we are going to reach
ultimate luminosities.~\cite{T1}

The new key element at photon colliders is a powerful laser system
which is used for e$\to \gamma$ conversion.  Lasers with required
flash energies and pulse duration already exist and are used in
several laboratories, the main problem here is the repetition rate.
Present technology would already allow the required laser systems to be 
built now, but it would be very expensive.~\cite{Perry} One  very
promising way to overcome this problem is discussed in this paper.
It is an optical cavity approach, which allows a considerable reduction of
the required  peak and average laser power.

As you know, in \EPEM\ collisions at linear colliders (LC), the beams
should be flat in order to restrict the beamstrahlung energy losses.
The typical beam sizes at the interaction point (IP) in the current
designs are about $\SX /\SY\ = (300-500)/(3-5)$ nm. Photon colliders
with the energies of several hundred GeV can work with practically
round beams with a radius of about 1--3 nm. Due to some technical
problems connected with the ``crab crossing'' and the ``big bend'' and
some increase of backgrounds due to a coherent pair creation obtaining
and operation with such small horizontal beam sizes at the IP is
problematic, but $\sigma_x \sim 10-15$ nm and $\sigma_y \sim 2$ nm is
quite a realistic goal.

 The main problem in achieving ultimate \GG\ luminosities is the generation of
electron beams with very small emittances both in the vertical and
horizontal planes. Damping rings can produce, in principle, the
required vertical emittance, but the horizontal emittance is larger
than  desired by two orders of magnitude.  Production of such low
emittances in both transverse directions is a very challenging task. Now
I see only one method to reach this goal, it is laser
cooling.~\cite{TSB1}$^,$\cite{Monter} The required laser system should
be much more powerful than that needed for e$\to \gamma$ conversion,
but it is not impossible that using the optical cavity scheme such
a system can already be built  now.  The problems in the laser cooling
and possible solutions are discussed in sect.~\ref{cool}.

The third group of problems is connected with transportation of low
emittance beams to the interaction point, collision and removal of
the disrupted beams without generation of additional backgrounds. 

\section{Lasers, optics}

\subsection{Requirements for the laser, wave length, flash energy}   
   
  Laser parameters important for this task are:  laser flash
  energy, duration of laser pulse, wave length and repetition rate.
  The required wave length follows from the kinematics of Compton
  scattering.~\cite{GKST83} In the conversion region a laser
  photon with the energy $\omega_0$ scatters at a small collision
  angle $\alpha_0$ on a high energy electron with the energy $ E_0$.
The maximum energy of scattered photons (in direction of electrons)
$$
\omega_m=\frac{x}{x+1}E_0; \;\;\;\;
x=\frac{4E_0\omega_0}{m^2c^4}
 \simeq 15.3\left[\frac{E_0}{\TEV}\right]
\left[\frac{\omega_0}{eV}\right] = 
 19\left[\frac{E_0}{\TEV}\right]
\left[\frac{\mu m}{\lambda}\right].
$$
For example: $E_0$ =250\,\, GeV, $\omega_0 =1.17$ eV
($\lambda=1.06$ \MKM) (Nd:Glass laser) $\Rightarrow$ $x=4.5$ and
$\omega/E_0 = 0.82$.  The energy of the backscattered photons grows
with increasing  $x$.  However, at $x > 4.8$ the high energy photons
are lost due to \EPEM\ creation in the collisions with laser
photons.~\cite{TEL90} The maximum conversion coefficient (effective)
at $x\sim 10$ is about 0.33 while at $x < 4.8$ it is about 0.65 (one
conversion length). The luminosity in the first case will be smaller
by a factor of 4. Detailed study of dependence of the maximum \GG\ 
luminosity and monochromaticity on $x$  can be found elesewhere.~\cite{TEL90}

In the laser focus at photon colliders the field is so strong
that multiphoton processes can take place, for example, the electron
can scatter simultaneously on several laser photons. It is preferable to 
work in a regime where these  effects are small enough,
because the shape of the photon spectrum is better. Sometimes strong
fields can be useful. Due to transverse motion of electrons in the
laser wave the effective electron mass is increased and the threshold
of \EPEM\ production  is shifted to the higher beam
energies, a factor of 1.5--2 is possible without special problems
``simply'' by adding a laser power.  For some tasks, such as
the energy scanning of the low mass Higgs, the luminosity spectrum should
be very sharp, that is only possible  when multiphoton effects
are small.

From all this it follows that an existing powerful Terawatt solid state
laser with the wave length about 1 \MKM\ can be used for photon colliders up
to c.m.s. energies about 1 TeV. For low energy colliders (for study of
the low mass Higgs, for instance), the doubling  of the
laser frequency may be useful, this can be done with high efficiency, 
about 45 \%.

In the calculation of the required flash energy one has to take into
account the natural ``diffraction'' emittance of the laser beam, the
maximum allowed value of the field strength (characterized by the
parameter $\xi^2 = (eB\hbar/m\omega_0 c)^2$) and the laser
spot size at the conversion point which should be larger than that of
the electron beam. In the scheme with crab crossing the electron beam
is tilted in respect to the direction of motion that creates an
additional effective transverse beam size $\sigma_x =
\sigma_z\alpha_c/2$.  The result of MC simulation of $k^2$
(proportional to the \GG\ luminosity) as a function of the flash energy
and parameter $\xi^2$ (in the center of the laser bunch) are shown in fig.
1 and 2.
\begin{figure}[!htb]
 
\vspace{-0.9cm}
 
\hspace*{0cm}\begin{minipage}[b]{0.45\linewidth}
\centering
\vspace*{-0.cm}
\hspace*{-1.cm} \epsfig{file=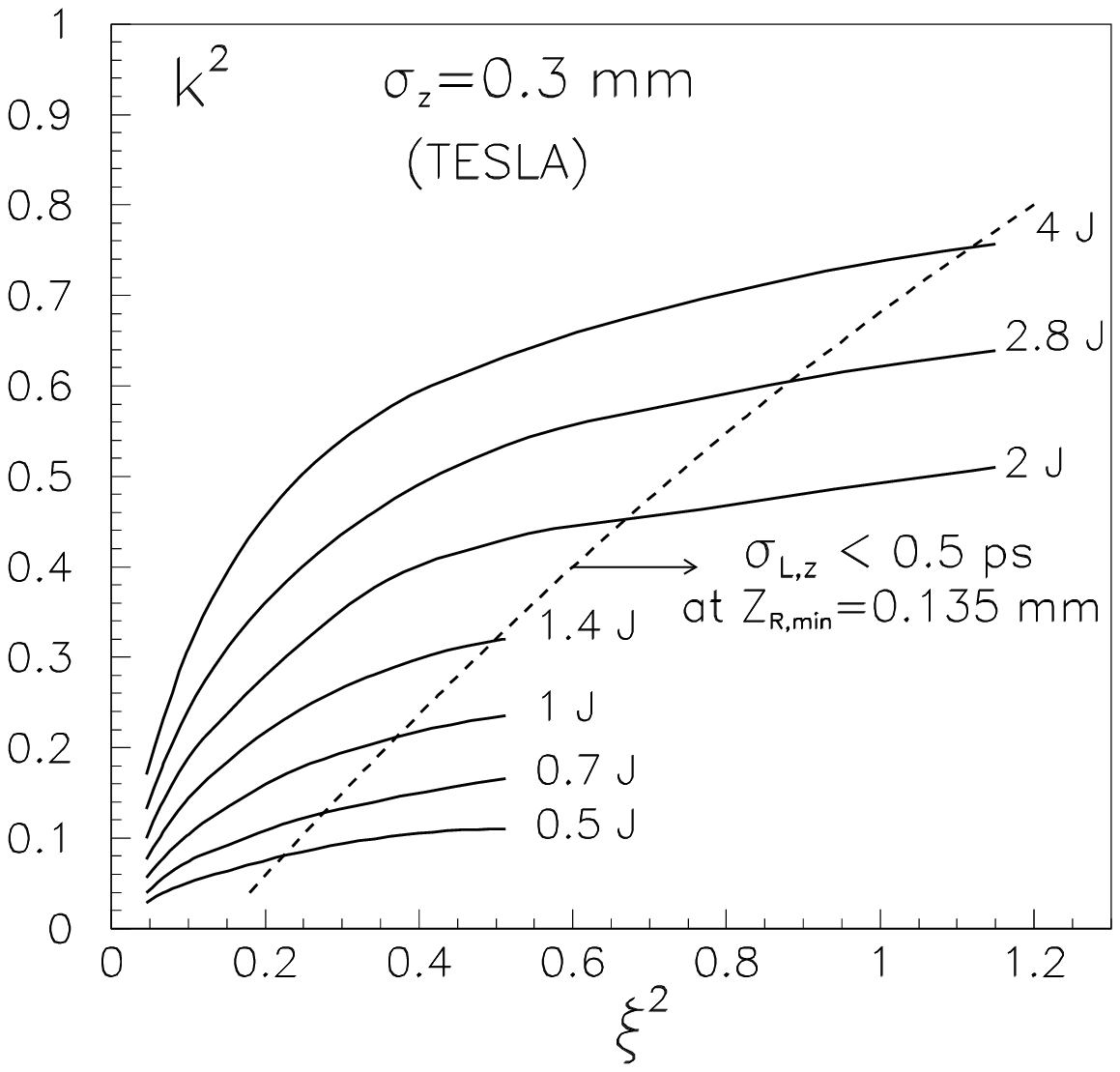,width=8.2cm,angle=0}
 
\vspace{-0.8cm}
\label{tes} 
\caption{The conversion probability for the various laser flash energies 
  and the values of the parameter $\xi^2$. Electron beams pass through
  the holes in the mirrors. See comments in the text.}
\end{minipage}%
\hspace*{1.1cm} \begin{minipage}[b]{0.45\linewidth} \centering
 
\vspace*{0cm}

\hspace*{-1.5cm} \epsfig{file=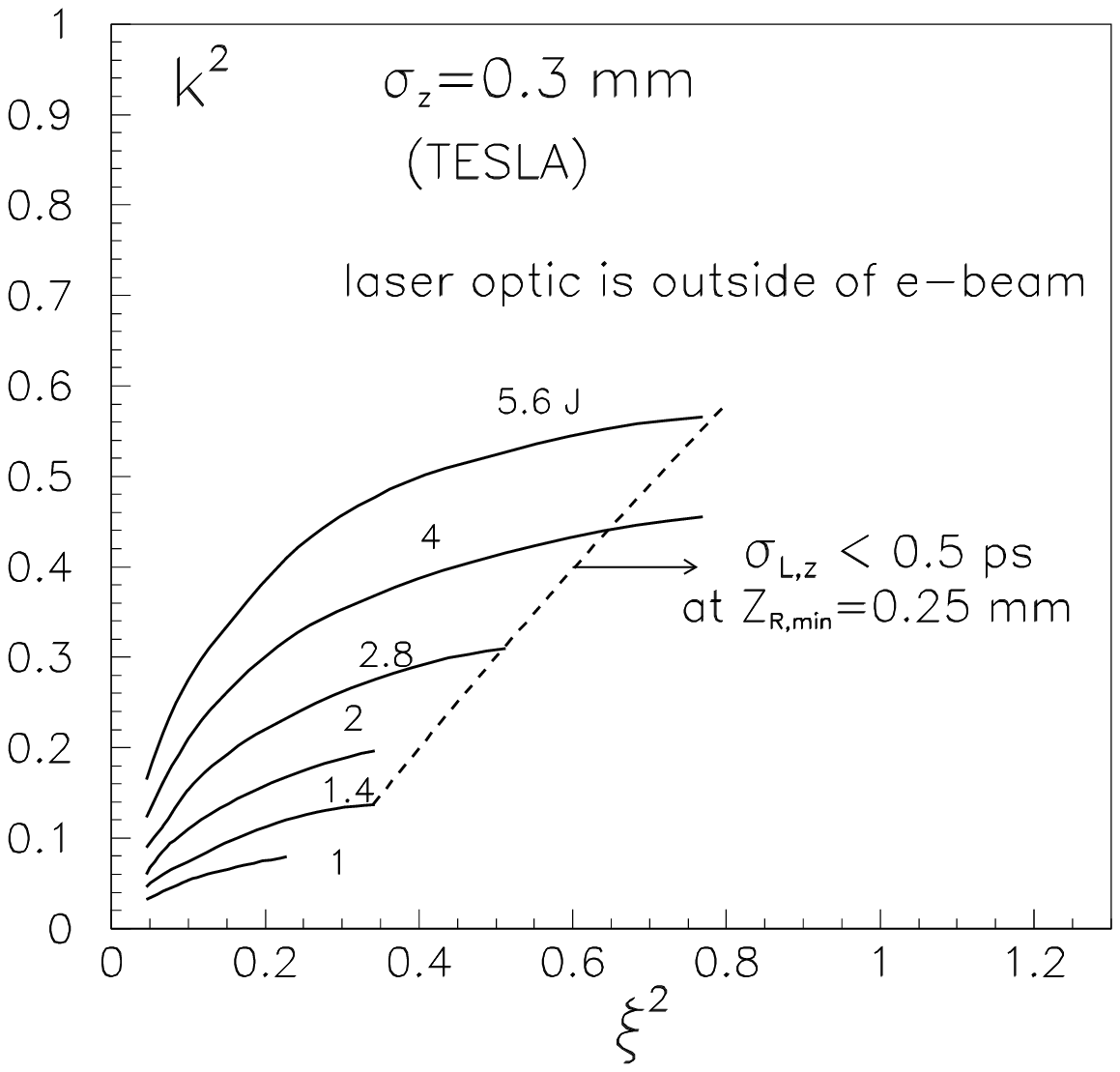,width=8.2cm,angle=0}
 
\vspace{-.7cm}
\label{tes2r}
\caption{Same as on fig.1, but the mirror system is situated outside 
the electron beam trajectories.\vspace{0.6cm}} 
\end{minipage}
\end{figure}

In summary: the required flash energy is about 3--5
Joules, that is quite reasonable. However, the LC have a repetition
rate of about 10-15 kHz, so the average power of the laser system should be
up to about 50 kW. \footnote{Though the average power in the one bunch
  train is higher, the cooling time (namely overheating of the
  crystals is the main problem) is longer than the time between
  trains, therefore we can speak about average power.} One  possible
scheme is a multi-laser system which combines  pulses into  one train
using Pockels cells.~\cite{NLC} However, such a system will be huge and
very expensive.~\cite{Perry}

\subsection{Multi-pass laser systems}

To overcome the ``repetition rate'' problem it is quite natural to
consider a laser system where one laser bunch is used for e$\to
\gamma$ conversion many times. Indeed, one Joule laser flash contains about
$10^{19}$ laser photons and only $10^{10}$ photons are knocked out in
the collision with one electron bunch. 

The simplest solution is to trap the laser pulse to some optical loop and
use it many times.~\cite{NLC} In such a system the laser pulse enters
via the film polarizer and then is trapped using Pockels cells and
polarization rotating plates.  Unfortunately, such a system will not
work with Terawatt laser pulses due to a self-focusing effect.

Fortunately, there is one way to ``create'' a powerful laser pulse in
the optical ``trap'' without any material inside. This very promising
technique is discussed below.

\subsection{Laser pulse stacking in an ``external'' optical cavity.} 

Shortly, the method is the following. Using the train of low energy
laser pulses one can create in the external passive cavity
(with one mirror having some small transparency) an optical pulse of
the same duration but with much higher energy (pulse stacking). This
pulse circulates many times in the cavity each time colliding with
electron bunches passing the center of the cavity.

The idea of pulse stacking is simple but not trivial and not well
known in the HEP community (and even to laser experts, though it is as old
as the Fabry-Perot interferometer). This method is used now in several
experiments on detection of gravitation waves. It was mentioned also
in NLC ZDR~\cite{NLC} though without analysis and further development.
   In my opinion, pulse stacking is very natural for photon colliders
and allows not only to build a relatively cheap laser system for
$e\to\gamma$ conversion but gives us the practical way for realization of
laser cooling, i.e. opens up the way to ultimate luminosities of photon
colliders. 

\begin{figure}[!htb]
\centering
\vspace*{0.2cm} 
\hspace*{-0.2cm} \epsfig{file=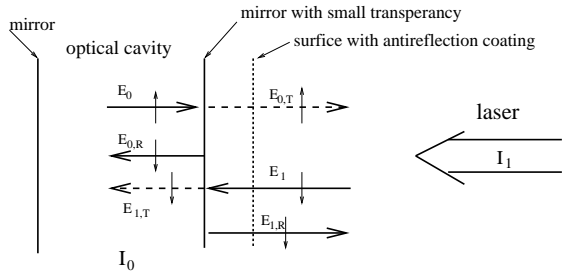,width=7.5cm,angle=0} 
\vspace*{-0.cm} 
\caption{Principle of pulse stacking in an external optical cavity.}
\vspace{2mm}
\label{cavity}
\vspace{-2mm}
\end{figure}

As this is very important for photon colliders, let me consider this
method in more detail. The principle of pulse stacking  is shown in
Fig.\ref{cavity}.
The secret consists in the following. There is a well known optical
theorem: at any surface, the reflection coefficients for light coming
from one and the other sides have opposite signs. In our case, this means
that light from the laser entering through semi-transparent mirror into
the cavity interferes with reflected light inside the cavity {\bf
constructively}, while the light leaking from the cavity interferes
with the reflected laser light {\bf destructively}. Namely, this fact produces
asymmetry between cavity and space outside the cavity!  

Let R be the reflection coefficient, T the transparency coefficient
and $\delta$ the passive losses in the right mirror. From the energy
conservation $R+T+\delta =1$. Let $E_1$ and $E_0$ be the amplitudes
of the laser field and the field inside the cavity. In equilibrium,
$E_0= E_{0,R} + E_{1,T}$. 
Taking into account that $E_{0,R}=E_0\sqrt{R}$, $E_{1,T}=E_1\sqrt{T}$ and 
$\sqrt{R}\sim 1-T/2-\delta/2$ for $R\approx 1$ we obtain
$E_0^2/E_1^2=4T/(T+\delta)^2.$ 
The maximum ratio of intensities is obtained at $T=\delta$, then 
$I_0/I_1=1/\delta \approx Q$,
where $Q$ is the quality factor of the optical cavity.  Even with two
metal mirrors inside the cavity, one can hope to get a gain factor of about
50--100; with multi-layer mirrors it can reach $10^5$. ILC(TESLA)
colliders have 120(2800) electron bunches in the train, so the factor
100(1000) would be perfect for our goal, but even the factor of ten
means a drastic reduction of the cost.

   Obtaining of high gains requires a very good stabilization of cavity
size: $\delta L \sim \lambda/4\pi Q$, laser wave length: $\delta
\lambda/\lambda \sim \lambda/4\pi QL$ and distance between the laser
and the cavity: $\delta s \sim\lambda/4\pi$. Otherwise, the  condition of
constructive interference will  not be fulfilled. Besides, the
frequency spectrum of the laser should coincide with the cavity modes,
that is automatically fulfilled when the ratio of the cavity length and
that of the laser oscillator is equal to an integer number 1, 2, 3... . 

For $\lambda = 1\;\mu m$ and $Q=100$, the stability of the cavity
length should be about $10^{-7}$ cm. In the LIGO experiment
 on detection of gravitational waves which uses 
similar techniques with $L\sim 4$ km and $Q\sim 10^5$ the expected
sensitivity is about $10^{-16}$ cm.  In comparison with this project
our goal seems to be very realistic.

      In HEP literature I have found only one reference on pulse
stacking of short pulses ($\sim 1$ ps) generated by FEL~\cite{HAAR}
with the wave length of 5 $\mu$m. They observed pulses in the cavity
with 70 times the energy of the incident FEL pulses, though no long
term stabilization was done.
   
\begin{figure}[!htb]
\centering
\vspace*{-0.0cm} 
\hspace*{-0.4cm} \epsfig{file=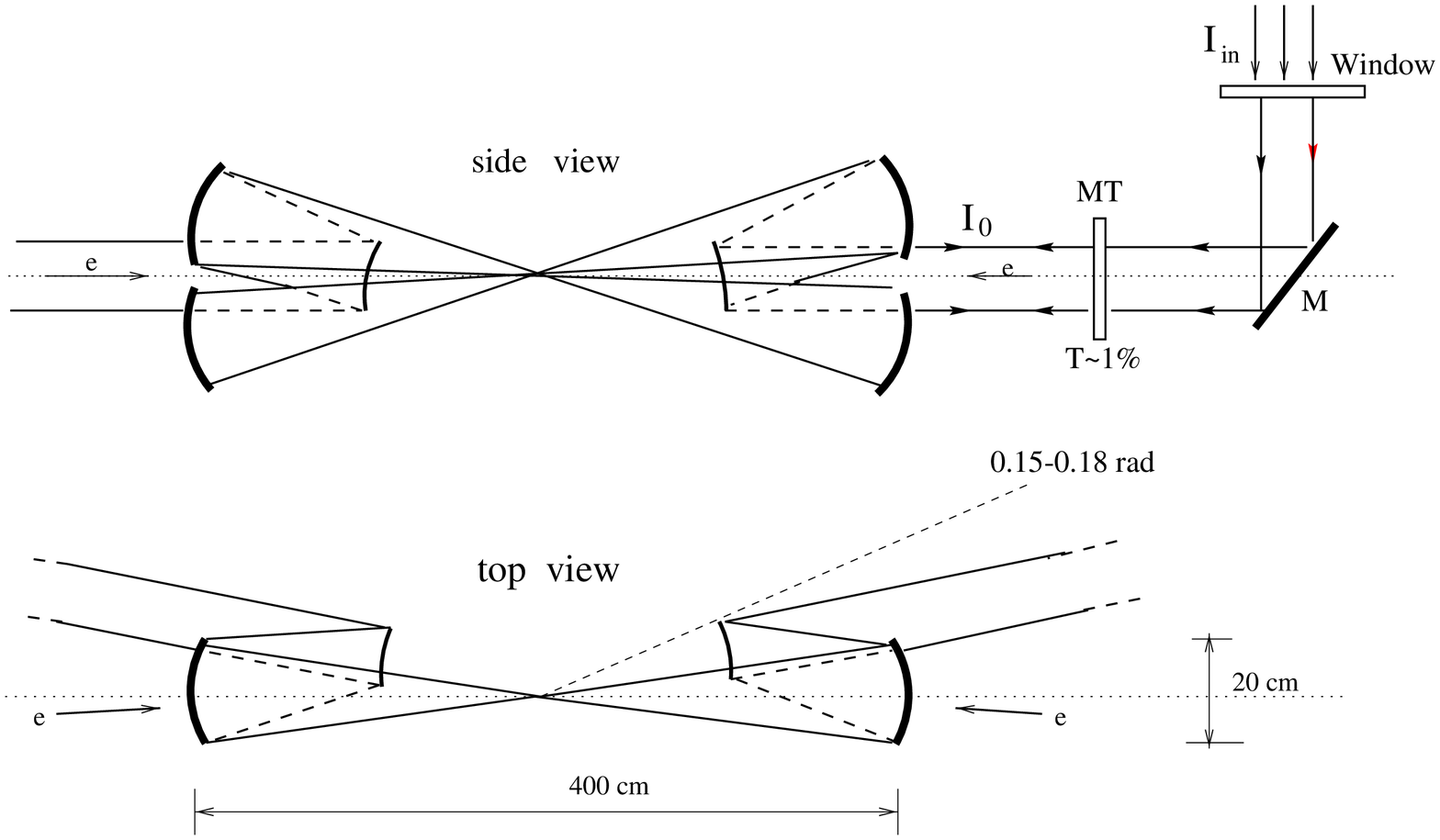,width=8cm,angle=0} 
\vspace*{-0.0cm} 
\caption{Possible scheme of optics at the IR.}
\vspace{0mm}
\label{optics}
\vspace{-6mm}
\end{figure}

Possible layout of the optics at the interaction region scheme is
shown in Fig.\ref{optics}. In this variant, there are two optical
cavities (one for each colliding electron beam) placed outside the
electron beams.  Another possible variant has only one cavity common
for both electron beams. In this case, it is also possible to arrange
two conversion points separated by the distance of several millimeters
(as it is required for photon colliders), though the distribution of
the field in the cavity is not completely stable in this case (though
it may be sufficient for not too large a Q and , it can be made stable in more complicated optical system). Also, mirrors should have
holes for electron beams (which does not change the Q factor of the
cavity too much). The previous variant is simpler though it requires
a factor of 2 higher flash energy.

\section{Laser cooling of electron beams \label{cool}}

The use of pulse stacking in the optical cavity makes the idea of
laser cooling~\cite{TSB1}$^,$\cite{Monter} very realistic, though the
required flash energy is one order higher than that required for $e
\to \gamma$ conversion.
In the method of laser cooling the electron beam at an energy of about 5
GeV (just after the damping ring and longitudinal compression) is collided
1--2 times with a powerful laser flash losing in each collision a
large fraction ($\sim 90$ \%) of its energy to the radiation (Compton
scattering), with re-acceleration between cooling sections. The physics
of the cooling process is almost the same as radiative cooling of
electrons in damping rings. However, here the process  takes
only 1 ps and the ultimate emittance is much lower than that in the damping
rings. This is because in the ``linear'' laser cooling there are no
bends which cause a  growth of the horizontal emittance. Also the intra-beam
scattering is not important due to a short ``damping'' time and  following
fast acceleration.
Considering a practical scheme for laser cooling we should take into
account many important practical aspects: 

$\bullet$ Radiation damage of the mirrors. X-ray radiation due to the
Compton scattering here is  many orders larger than the radiation level
at the same angles in the $\gamma \to e$ conversion point.  It is so
because a) the electron energies are lower and b) each electron undergoes
about one hundred Compton scattering.  At $\vartheta \gg 1/\gamma$ and $x
\ll 1$ ($x$ is defined in sect.2) the energy of the Compton scattered
photons $\omega = 4\omega_0/\vartheta^2$ and does not depend on the
electron energy.~\cite{GKST83} However, at the lower beam energies the
spectrum is softer ($\omega_{max} = 4\omega_0\gamma^2)$ and more
photons (per one Compton scattering) have large angles. Simple
calculations show that the number of photons/per 
electron emitted on the angle $\vartheta$ during the cooling of electrons
from some large energy to the energy $E_{min}$ is 
$$
dn/d\Omega =mc^2/4\pi\omega_0\gamma_{min}^3\vartheta^4.
$$

The total energy  hitting the mirrors/cm$^2$/sec is 
$$
dP/dS=mc^2N\nu/\pi\gamma_{min}^3\vartheta^6 L^2,
$$
where $L$ is the distance between the collision (cooling) point
(CP) and the focusing mirrors, N and $\nu$ are the number of electrons in
the bunch and the collision rate. One can see a strong dependence of X-ray
background on $\gamma_{min}$ and $\vartheta$. During the cooling the
electron beam loses almost all its energy to photons. For $E_0=5$ GeV,
$N=2\times 10^{10}$, $\nu = 15$ kHz the total energy losses are about
200 kW, fortunately the flux decreases rapidly with increasing the angle. At
$\vartheta$ = 30 mrad and $L=5$ m the power density $dP/dS \sim
10^{-5}$ W/cm$^2$ and X-ray photons have an energies of about 4 keV (for 1
\MKM\ laser wave length). My estimations shows that rescattering of photons on
the quads can give a comparable background. 

I have describing this item in detail because for laser cooling the
required flash energy is very high and to reach the goal we need very
high reflectivity of the mirrors in the optical cavity. For TESLA with
3000 bunches in a train it would be nice to have mirrors with
$R>0.999$. Such values of R are not a problem for dielectric mirrors,
however the radiation damage may cause problems, better to avoid this
problem.

$\bullet$ Laser spot size should be several times larger than that of
the focused electron beam to avoid an additional energy spread of
the cooled electrons.

$\bullet$ The cooled electron beam at the energy E=500-1000 GeV has an
energy spread of $\sigma_E/E \sim 15$ \% at the point where the
$\beta$- function is small ($\sim 1-5$ mm). Matching
this beam with the accelerator is not a simple problem and requires
special insertions for chromaticity correction. A similar
problem exists for the final focus at linear colliders, it has been
solved and tested at the FFTB at SLAC. Here the factor
$(F/\beta)\sigma_E/E$ characterizing the chromaticity problem is
smaller and the beam energy is 500 times smaller, so one can hope that
it will be no problem.

$\bullet$ The parameter $\xi^2$ (defined above) should be small enough
($\le$ 1) to keep the minimum attainable emittance, depolarization and the
energy spread small enough.\cite{TSB1}$^,$\cite{Monter} This is
impossible with one laser (with required flash energy) without
additional "stretching" of the cooling region along the beam line.
The simplest way to do this is to focus several lasers
at different points along the beam axis.

\begin{figure}[!htb]
\centering
\vspace*{-0.0cm} 
\hspace*{-0.4cm} \epsfig{file=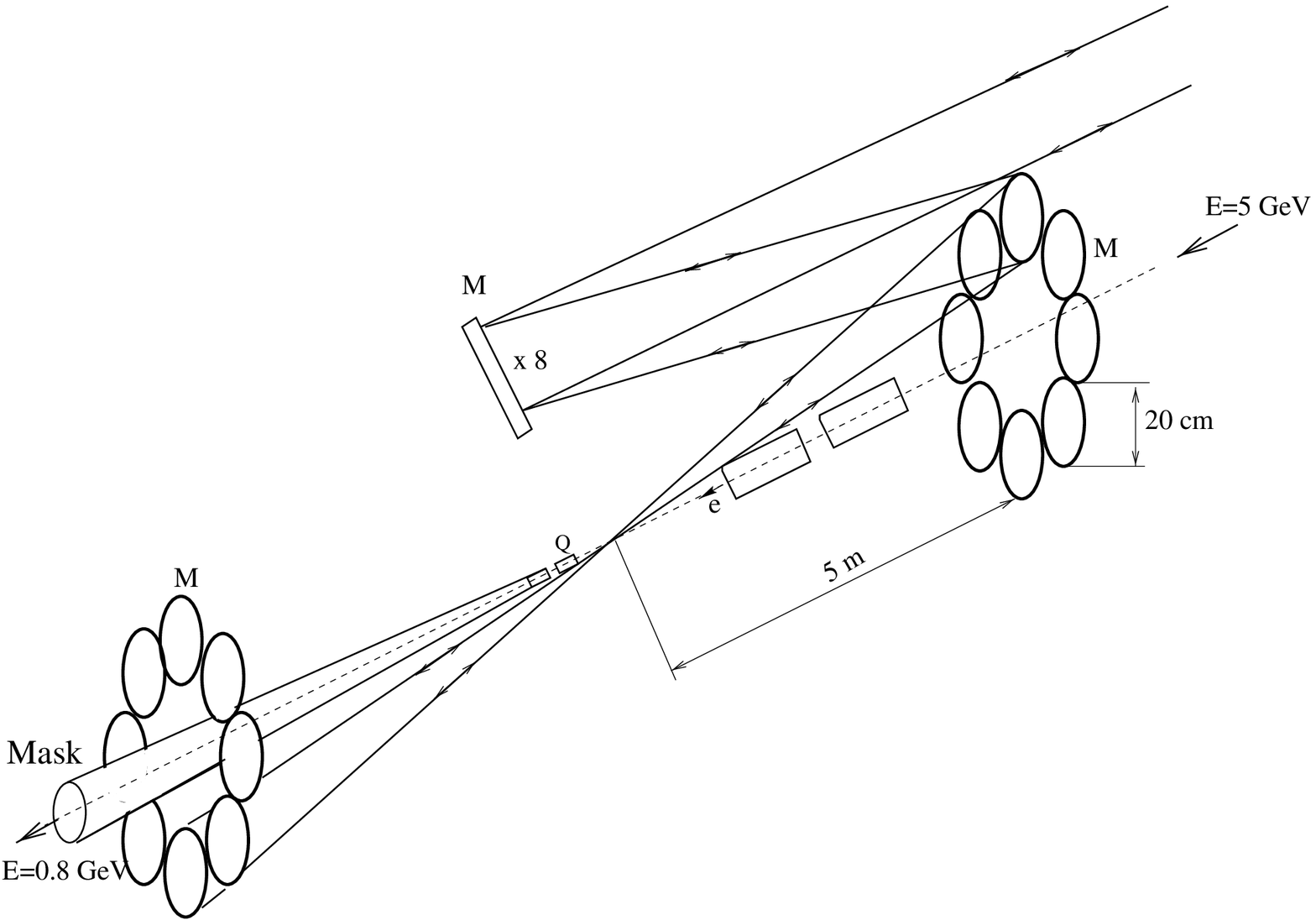,width=8cm,angle=0} 
\vspace*{-0.0cm} 
\caption{Possible scheme of laser cooling.}
\vspace{0mm}
\label{cooling}
\vspace{-0mm}
\end{figure}

The possible optical scheme for the TESLA project is shown in
fig.\ref{cooling} (only the final focusing
mirrors are shown). The system consist of 8 independent identical
optical cavities focusing the laser beams to the points distibuted
along the beam direction on the length $\Delta z \sim 2$ mm. The
length of the cavity (the distance between the ``left'' mirror and an
entrance semi-transparent mirror (not shown)) is equal to half the
distance between the electron bunches in the train, 50 m for TESLA).
The large enough angle between the edges of the mirrors and the beam
axis (30 mrad) makes X-ray flux rather small (see the estimation
above).  Also this clear angle allows the final quads to be placed at
a distance about 50 cm (from the side of the cooled beam), much closer
than the focusing mirrors. Smaller focal distance makes the problem of
chromaticity correction easier.

The maximum distance from the CP to the mirrors is determined only by
the mirror size, the diameter of 20 cm seems reasonable, which gives
$L=5$ m.  The laser spot size at the CP is 7.5 \MKM, at least 3 times
larger than the horizontal electron beam size with $\beta_x <$ 5 mm.
The circulating flash energy in each cavity is 25 J and 200 J in the
whole system, not small. The average power circulating inside the system is
$200\times 15$ kHz = 3 MW!  However, if the Q factor of the cavities
is about 1000--3000 (3000 bunches in the electron train at TESLA), the
required laser power is only 1--3 kW, or 0.15--0.4 kW/per each laser,
that is already reasonable.

What about damage to the mirrors by such powerful laser light?  The
maximum laser flash energy/cm$^2$ on the mirrors is 0.13 J/cm$^2$
(0.7-2 has been achieved for 1 ps pulses~\cite{NLC}), the average
power/cm$^2$ is 2 kW/cm$^2$ (there are systems with $>5$ kW/cm$^2$
working long time~\cite{NLC}). The average power inside one train
($\Delta t = 1$ msec) is 200 times higher (400 J/1 msec), but from the
same ref.\cite{NLC} is known that 100 J for a time of 100 ns is OK,
and extrapolating as $\sqrt{t}$ (thermoconductivity) one can expect
the limit of about 10 kJ for 1 msec, much larger that expected in our
case.  Note, here we are speaking about circulating, not absorbed
energy.  So, all power densities are below the known limits, this all
depends, of course, on specific choice of mirrors.

At last, the main numbers. After one stage of such a cooling system
the normalized emittance is decreased by a factor of 6. The ultimate
normalized emittance (after several cooling sections) is proportional
to the $\beta$-function at the CP, at $\beta_{x,y}=1$ mm it is about
$2\times10^{-9}$ m rad, smaller than can be produced by the TESLA
damping ring by a factor of 5000(15) in x(y) directions. From this
point of view such a small $\beta_x$ is not necessary, but it should be
small enough ($< 5$ mm to have a small electron spot size in the
cooling region.  The first stage of cooling will be the most efficient
because  the beam is cooled in both horizontal and vertical directions (far
from the limits). Besides after decreasing the horizontal emittance the
$\beta$- function at the LC final focus can be made as small as
possible, $\sim \sigma_z.$ All together this can give a factor of ten in
the luminosity.

Having no space for discussion of accelerator aspects in this paper I
would like to note only that all systems of the LC should allow beam
emittances to be reached which are lower than are necessary for \EPEM\ 
collisions (see the intoduction). Many technical decisions should be done
before the beginning of construction works.

\section{Conclusion}
  Photon colliders is a very inspiring new field of high energy physics and
I invite you to take part in this venture.

\end{document}